\documentclass{INTERSPEECH2023}
\usepackage{amsmath,graphicx}
\usepackage{threeparttable}
\usepackage{float}
\usepackage{subfigure}
\usepackage{graphicx}
\usepackage{multirow}
\usepackage{booktabs}
\usepackage{tabularx}
\usepackage{hyperref}
\usepackage{amssymb}

\title{DCCRN-KWS: An Audio Bias Based Model for Noise Robust Small-Footprint Keyword Spotting}

\name{Shubo Lv$^{1,2}$$^{*}$\thanks{* Equal contribution. $\dagger$ Lei Xie is the corresponding author.}, Xiong Wang$^{2}$$^{*}$, Sining Sun$^{1}$, Long Ma$^{2}$, Lei Xie$^{1\dagger}$}
\address{$^1$Audio, Speech and Language Processing Group (ASLP@NPU) \\
$^2$Tencent Technology Co., Ltd, Beijing, China}
\email{shblv@npu-aslp.org, chnxwang@tencent.com, ssning2013@gmail.com,  malonema@tencent.com, lxie@nwpu.edu.cn}

\begin{document}

\maketitle
 
\begin{abstract}
Real-world complex acoustic environments especially the ones with a low signal-to-noise ratio (SNR) will bring tremendous challenges to a keyword spotting (KWS) system. Inspired by the recent advances of neural speech enhancement and context bias in speech recognition, we propose a robust audio context bias based DCCRN-KWS model to address this challenge. We form the whole architecture as a multi-task learning framework for both denoising and keyword spotting, where the DCCRN encoder is connected with the KWS model. Helped with the denoising task, we further introduce an audio context bias module to leverage the real keyword samples and bias the network to better discriminate keywords in noisy conditions. Feature merge and complex context linear modules are also introduced to strengthen such discrimination and to effectively leverage
contextual information respectively. Experiments on an internal challenging dataset and the HIMIYA public dataset show that DCCRN-KWS is superior in performance, while the ablation study demonstrates the good design of the whole model.
\end{abstract}
\vspace{-0.2cm}
\section{Introduction}
\label{sec:intro}
\vspace{-0.1cm}
Keyword Spotting (KWS) is a task to detect whether the input speech signal contains preset keywords. Accordingly, as a typical KWS task, wake-up word (WuW) detection is the first step in voice interaction between the user and smart devices~\cite{miller2007rapid}. However, there is usually a certain interaction distance between the user and the device, whereas speech signal decay, environmental noise, and room reverberation will seriously affect the performance of the system.%

To improve speech quality, a signal enhancement front-end is usually adopted. When multi-channel signals can be collected from a microphone array, beamforming or multi-channel signal enhancement technologies can be adopted. Traditional signal processing based speech enhancement methods usually apply a spectral suppression gain (or filter) to the noisy signal under the statistical signal processing theory~\cite{wang2006computational}. With the help of deep learning (DL), speech enhancement has been recently formulated as a supervised learning problem and has become the mainstream because of their strong noise reduction abilities (especially for non-stationary noise) learned from simulated clean-noisy speech pairs~\cite{williamson2015complex, tan2018convolutional}. More recently, advanced network structures such as SDD-Net~\cite{li2021simultaneous} and DCCRN~\cite{hu2020dccrn}, which explicitly model the complex spectrum of speech with particularly designed optimization metrics, have shown outstanding performance in recent noise suppression challenges~\cite{reddy2020interspeech, reddy2021interspeech}.

The advances of neural speech enhancement have triggered its downstream applications in speech recognition~\cite{o2021conformer}. The front-end speech enhancement module can be optimized independently in prior~\cite{nian2022time} 
or jointly with the acoustic model~\cite{li2021espnet} 
to improve the noise robustness of ASR. In~\cite{kong2021multi}, the architecture of deep complex Unet (DCUnet)~\cite{kong2021multi} is incorporated with a multichannel acoustic model by multi-task learning. 
The advantages of front- and back-end integration can be directly applied to KWS which specifically aims to recognize predefined keywords~\cite{gu2020efficient}. However, specifically for the KWS task, while considering the integration, we can better leverage the constrained linguistic and acoustic information as keywords are fixed or limited in number. In this direction, a neural network based \textit{text-dependent} speech enhancement (TDSE) method was proposed in~\cite{yu2018text} for recovering the original clean speech signal of a specific keyword. In the open-vocabulary KWS approach in~\cite{shin2022learning}, an attention-based text-audio cross-modal matching approach was proposed, where a denoising loss for the acoustic embedding network is specifically used to improve noise robustness.


Aiming to improve the KWS performance under noisy conditions and make full use of the prior information of keywords, we propose a novel \textit{audio context bias} based front-end and back-end integrated KWS model. Inspired by~\cite{kong2021multi}, we cascade the DCCRN encoder with a dilated temporal convolution (DTC) based ~\cite{hou2021npu} KWS model, resulting in the DCCRN-KWS model learned under multi-task learning framework. To further bias the model to learn specific keywords, inspired by the advances in context bias in ASR ~\cite{pundak2018deep}, we further introduce an \textit{audio bias encoder} which extracts specific keyword embedding from keyword audio samples. The keyword embedding, subsequently concatenated with the DCCRN encoder output, is fed into the KWS module. Based on the observation that each DCCRN encoder layer output has clear discrimination between keywords and non-keywords, we further introduce a \textit{feature merge} module to aggregate the outputs from each layer to further strengthen such discrimination. Finally, we propose a \textit{complex context linear} module to better integrate the bias embedding with the DCCRN encoder output, improving KWS performance further and avoiding model complexity explosion. Experiments on two challenging datasets clearly show the advantages of the proposed approach with superior performance and reasonable complexity. Ablation studies further demonstrate the good design of different modules.

\vspace{-0.3cm}

\section{Audio Context Bias for DCCRN-KWS}
\label{sec:pagestyle}
\vspace{-0.2cm}

In this section, we first introduce our back-end KWS module and then detail the proposed DCCRN-KWS architecture, followed by the design of audio context bias module, feature merge module and complex context linear module respectively.
\vspace{-0.3cm}
\subsection{Architecture of KWS module}
\begin{figure}[h]
\vspace{-5pt}
\centering
\includegraphics[width=1\linewidth]{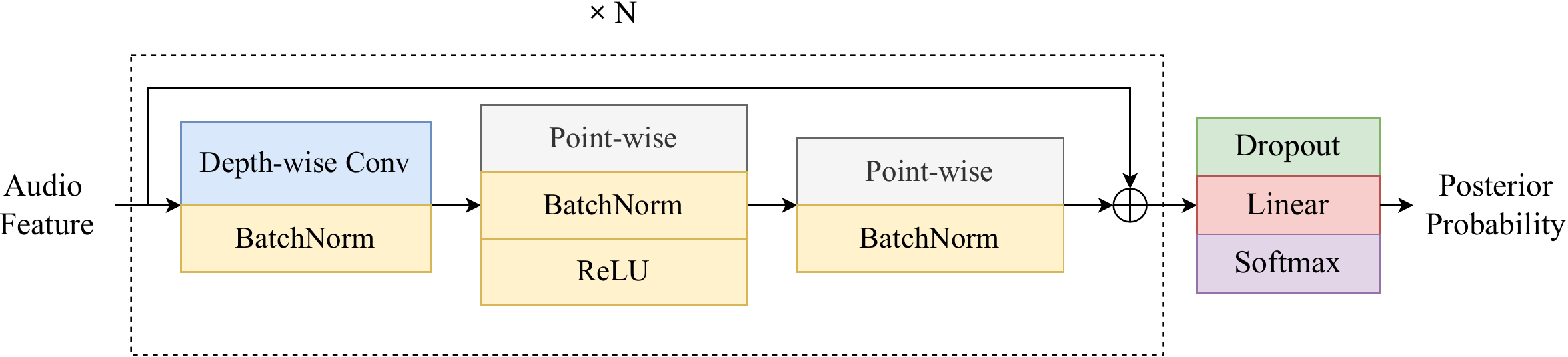}
\vspace{-0.6cm}
\caption{Architecture of KWS module.} 
\label{fig:kws}
\vspace{-15pt}
\end{figure}

\vspace{-0.1cm}
As shown in Fig.~\ref{fig:kws}, our KWS module consists of multiple Dilated Temporal Convolutional (DTC) blocks, which has achieved impressive performance recently~\cite{hou2021npu}. In detail, a dilated depthwise 1D-convolution layer is first used to obtain temporal context, followed by two layers of pointwise convolution to integrate the latent features from different channels. Finally, we use a fully connected (FC) layer with softmax function to estimate the posterior probability of the keyword.
\vspace{-0.3cm}
\vspace{-0.1cm}

\subsection{Network structure of DCCRN-KWS} 
\vspace{-0.1cm}
We choose DCCRN~\cite{hu2020dccrn} as our front-end speech enhancement model. Shaped with an encoder-decoder Unet structure, a deep complex convolution recurrent network was updated from the CRN network by introducing complex convolution and LSTM, leading to superior performance in noise suppression. A straightforward way to unify the DCCRN model with the KWS model is to directly compose them in a cascaded structure. In this way,  we expect the DCCRN front-end to learn
the latent representation that benefits KWS without signal supervision on enhancement and fully depends on the optimization of the KWS back-end.
However, besides the increased complexity of the whole model, previous work~\cite{kong2021multi} has shown that this direct composition framework is not as good as the multi-task learning (MTL) framework with two explicit optimization targets. Therefore, we shape our architecture in Fig.~2 in an MTL manner~\cite{kong2021multi}, where the main task is keyword spotting and the auxiliary task is speech enhancement. Specifically, we make the DCCRN encoder shared by the enhancement decoder and the KWS network in order to extract the latent representation of the input signals that can benefit both tasks. The whole 
architecture is optimized during training and the speech enhancement part is simply discarded during inference. Based on this architecture, we introduce several important modules, including audio context bias, feature merge and complex context linear, detailed in the following subsections.

\begin{figure}[t]
\centering
\includegraphics[width=1\linewidth]{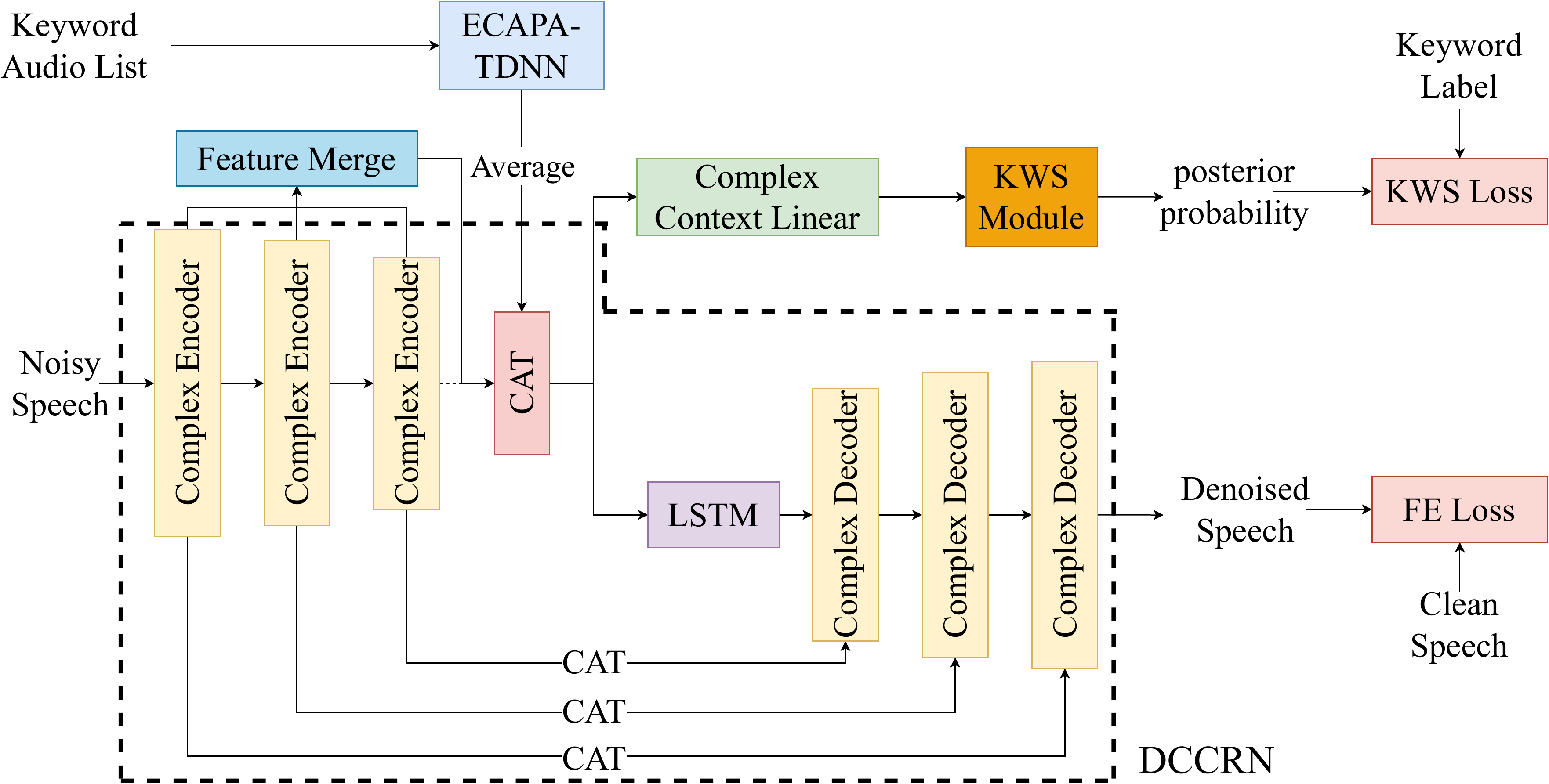}
\vspace{-0.3cm}
\caption{Audio Context Bias based DCCRN-KWS architecture.}
\vspace{-0.1cm}
\label{fig:all}
\vspace{-15pt}
\end{figure}
\vspace{-0.2cm}
\subsection{Audio Context Bias}
\vspace{-0.1cm}

Inspired by the context bias strategy in ASR~\cite{pundak2018deep}, we introduce a novel \textit{audio context bias} module into our DCCRN-KWS model. Different from extracting text embedding in~\cite{shin2022learning}, we apply an audio encoder to extract keyword audio embedding directly. The main reason is twofold. First, measuring audio-audio similarity is more straightforward than audio-text similarity; text bias is more flexible for customizing keywords but it is not the purpose of this paper. Second, such audio bias can better leverage real audio samples, further helping with the speech enhancement task to improve robustness in keyword discrimination in noisy conditions. 

In detail, we first make a keyword audio list that is selected from the keyword corpus. The list can be fixed (once chosen from the corpus, it is not changed anymore) or varied (randomly selected  from the corpus every time). Then we feed the keyword samples on the list to an embedding extractor to extract the bias embedding.  Here we adopt ECAPA-TDNN~\cite{desplanques2020ecapa} as the embedding extractor because this network shows superior ability in information aggregation from short audio clips and leads to SOTA performance in speaker verification. Finally, we average all the bias embedding vectors on the list and concatenate the average embedding with the last layer output of the DCCRN encoder. The concatenated final embedding is used as the input of the KWS module. Please note that when the keyword audio list is fixed, the keyword bias embedding is fixed accordingly during the inference time. Therefore, in this case, the additional complexity of the audio context bias module is negligible at run-time as the embedding extractor is not in use. In our case, the dimension of the keyword embedding is set to 192 and ECAPA-TDNN is implemented by SpeechBrain toolkit~\cite{speechbrain}. The ECAPA-TDNN-based embedding extractor is jointly trained with the whole DCCRN-KWS structure.
\vspace{-0.3cm}
\subsection{Feature Merge}
\begin{figure}[h]
\vspace{-10pt}
\centering
\includegraphics[width=1.0\linewidth]{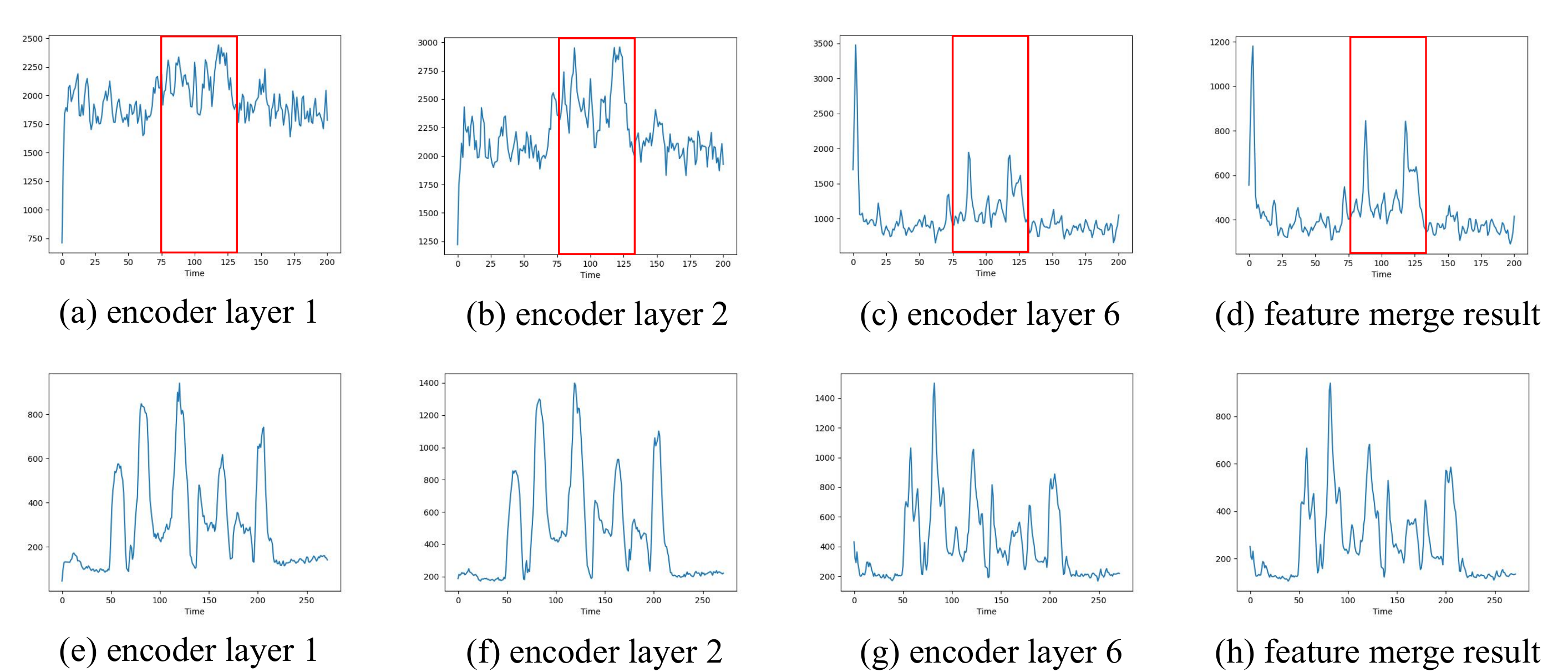}
\vspace{-0.4cm}
\caption{The energy of different encoder layers and the feature merge output for a keyword sample (up row) and a non-keyword sample (bottom row).} 
\label{fig:feature_merge}
\vspace{-15pt}
\end{figure}

\vspace{-0.1cm}
As we adopt the DCCRN encoder output as the feature for KWS, the purpose of each encoder layer should be analyzed. To this end, we visualize different encoder layer's output via frame level \textit{energy} ${m_t}$:
\begin{equation}\label{frame-level energe}
\vspace{-0.3cm}
\begin{cases}
m_{t} &= \sum_{c} \sum_{f} M_{c,f,t} \\
M_{c,f,t} &= \sqrt{\Re{(E)}_{c,f,t}^{2} + \Im{(E)}_{c,f,t}^{2}} 

\end{cases}
\end{equation}
where ${E_{c,f,t} \in \mathbb{R}^{C\times F\times T}}$ denotes each encoder layer and ${C, F, T}$ denote the channel number, frequency bins, and time frames respectively. $\Re{(E)}$ and $\Im{(E)}$ represent the real and imaginary parts of the encoder output respectively. Using a keyword sample and a non-keyword sample as input, we visualize the layer-wise energy in Fig~\ref{fig:feature_merge}.
We can see that the keyword part shows relatively higher energy as compared with the rest part, and this phenomenon is more salient for higher encoder layers. In contrast, this phenomenon is not the case for the non-keyword sample, where the energy distribution has a similar pattern for different layers. Whereas, we can conclude that the DCCRN encoder aims to enhance the \textit{energy} of the keyword part which can subsequently benefit the KWS module to better distinguish the keyword from other audio parts.

Motivated by this phenomenon, we develop a \textit{feature merge} module to increase the discrimination between keywords and others. Specifically, we first initialize a group of feature merge ratio $w_{i}$, where the last layer's ratio is 1 and other layers are 0. Then as the dimension of the previous encoder layer is twice than that of the final layer, we downsample the output of these layers. Finally, we weighted-average the outputs of the encoder layers to obtain the feature merge output $E'$. The process can be summarized as
\begin{equation}\label{feature merge}
\vspace{-0.3cm}
E' = \sum_{i}w_{i}\ast \text{downsample}(E_{i})  / \sum_{i}|w_{i}|
\end{equation}
where $w_{i}$ is learnable weights during training and fixed during inference. From Fig~\ref{fig:feature_merge} (d) and (h), we can see that the keyword segment becomes more salient after the feature merge while the non-keyword is almost unchanged.
\vspace{-0.3cm}
\subsection{Complex Context Linear}
\vspace{-0.1cm}
A common strategy of combining the bias embedding and encoder output is simply concatenating them together and subsequently going through a fully connected layer. As we know, contextual information is essential for the KWS task. To better model context, another strategy is concatenating several history frames of temporal features of the encoder and the bias embedding as the input of the fully connected layer. However, this simple concatenation will lead to a computational complexity explosion.

\begin{figure}[h]
\centering
\includegraphics[width=1\linewidth]{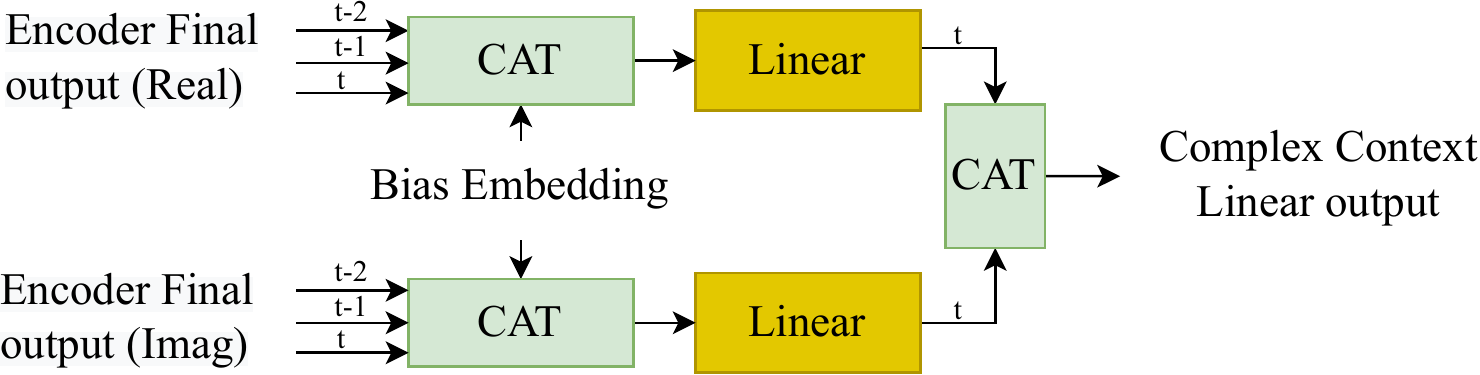}
\caption{Complex context linear module.} 
\label{fig:cplx_context_linear}
\vspace{-15pt}
\end{figure}

In order to better take the encoder context into account and avoid the complexity explosion, we specifically modify the fully connected layer shown in Fig~\ref{fig:cplx_context_linear}. We first split the encoder's output into real/imag parts separately and then concatenate the real/imag parts with bias embedding respectively. 
Finally, the context features of current ($t$) and previous frames ($t-1,t-2$) are combined together as the fully connected layer's input. 
Similar to the group convolution, the complexity is almost halved as compared with simple context concatenation.

\vspace{-0.2cm}
\subsection{Loss Function}
\vspace{-0.1cm}
We adopt the typical time-domain SI-SNR~\cite{luo2019conv} loss function for the speech enhancement task, while binary cross entropy (BCE) loss is used as the KWS loss. The final loss can be formulated as
\begin{equation}
\vspace{-0.1cm}
     \setlength{\arraycolsep}{0.3pt}
     \begin{cases}
     
     \mathcal{L}_{\textbf{BCE}} &= -y_{i}*ln(y_{i})-(1-y*)ln(1-y_{i}) \\
     
     \mathcal{L} &= \mathcal{L}_{\textbf{SI-SNR}} + \mathcal{L}_{\textbf{BCE}}
     \end{cases}
\vspace{-0.1cm}
\end{equation}
where ${y_{i}=H(x_{i};\theta )\in \{0,1\}}$ is the posterior predicted by the KWS model $H$ with parameter $\theta$, ${y*_{i}\in \{0,1\}}$ is the ground-truth class label for frame $i$.


\vspace{-0.2cm}

\section{Experiment}
\vspace{-0.2cm}
\label{sec:typestyle}
\subsection{Dataset}
\vspace{-0.1cm}
We first conduct ablation experiments to prove the effectiveness of each proposed sub-modules on a large internal training set and the task is to detect the keyword ``\textit{ding1-dang2-ding1-dang2}" in Mandarin. Specifically, a 158-hour keyword-specific speech dataset is used as positive training data, and a 139-hour speech dataset collected from Youtube is served as negative training data. The test is recorded in far-field noisy conditions, where the number of positive and negative samples are 239 and 12,480 respectively with SNR ranging from -6 to -2 dB.

Besides the internal task, we further validate our approach on the publicly available HIMIA dataset~\cite{qin2020hi} recorded from a high-fidelity microphone with spoken keywords ``\textit{ni2-hao3-mi1-ya4}" in Mandarin. The training set contains 10 hours of keyword samples from 254 speakers and the test set is composed of 
3,520 keyword samples from another 44 speakers. As speaker labels are known, speaker-dependent audio lists for keyword bias can be further studied. In this task, we adopt 90 hours of negative samples selected from the librivox speech corpus. 
For testing, besides the above positive testing samples, another 3,600 negative samples are randomly selected from librivox, in total 7,120 audio samples. 

During the training for both tasks, the training data are generated on the fly with SNR ranging from 0 to 15 dB and the image method is used to simulate RIR with RT60 ranging from 0.05s to 0.95s. For the internal set, the noises are selected from both an internal noise set and freesound,
174 hours in total. As for the HIMIYA public set, 56 hours of noise are selected from freesound. During training, we randomly add 1 to 4 types of noise to each speech utterance. Particularly, the speakers and noises have no overlap between training and testing. Furthermore, particularly for the HIMIYA set, we randomly clip the negative examples to short clips with a duration of 1 to 3 s for both training and testing in order to avoid the model overfitting to a specific time duration since the length of positive examples ranges from 1 to 3 s. To test our model's performance on different SNR conditions, we add noise to the HIMIYA testing samples under SNR of -5, 0, and 5 dB respectively.

\vspace{-0.3cm}
\subsection{Training setup and baselines}
\vspace{-0.1cm}
For the proposed model, the window length and frame shift are 25ms and 10ms respectively while the STFT length is 512. Our model is a causal one without considering future frames. For both datasets, all models are trained for 17,500 iterations. And we use 8 NVIDIA V100 32GB GPUs for the internal dataset and 1 for the public one. Adam optimizer is used with a Noam-based scheduler~\cite{dong2018speech}, where the Noam factor is 5 and the warm-up step is 1,000. During training, the number of the audio bias list is set to 50 and the input dimensions of the KWS module are 128. Furthermore, the fully connected layer between the final encoder output and the KWS module for DCCRN-KWS (w/o audio context bias), DCCRN-KWS (w/o complex context linear) and DCCRN-KWS (w complex context linear) are 1024 $\times$ 128, 1216 $\times$ 128, 608 $\times$ 64 $\times$ 3 separately. For the complex context linear module, the context number is set to 3 (the previous two frames with the current frame). The model details are described as follows. 
\begin{itemize}
    \item \textbf{KWS module}: The number of DTC layers is 16, and the kernel size and dilation are 5 and 1-2-4-8-1-2-4-8-1 respectively. To achieve streaming inference, all convolutions are causal ones. 
    \item \textbf{DCCRN}: The number of channels for the DCCRN is \{16,32,64,128,256,256\}, and the convolution kernel and step are set to (5,2) and (2,1) respectively. Two 256-dimensional LSTM layers are adopted with a 256-dimensional project layer. Each encoder/decoder module handles the current frame and one previous frame. 
\end{itemize}

During training, we select up to 10 frames around the last frame of a keyword audio clip as positive training samples and assign 1 to them, while other frames in the keyword audio clip are discarded as ambiguous and are not used in training. For negative training utterances, all frames are regarded as negative training samples and assigned to 0 accordingly.

\vspace{-0.3cm}
\subsection{Experimental results and discussion}
\label{sec:majhead}
\vspace{-0.1cm}
\subsubsection{Results on internal dataset}
\begin{figure}[t]
\centering
\vspace{-0.3cm}
\includegraphics[width=0.8\linewidth]{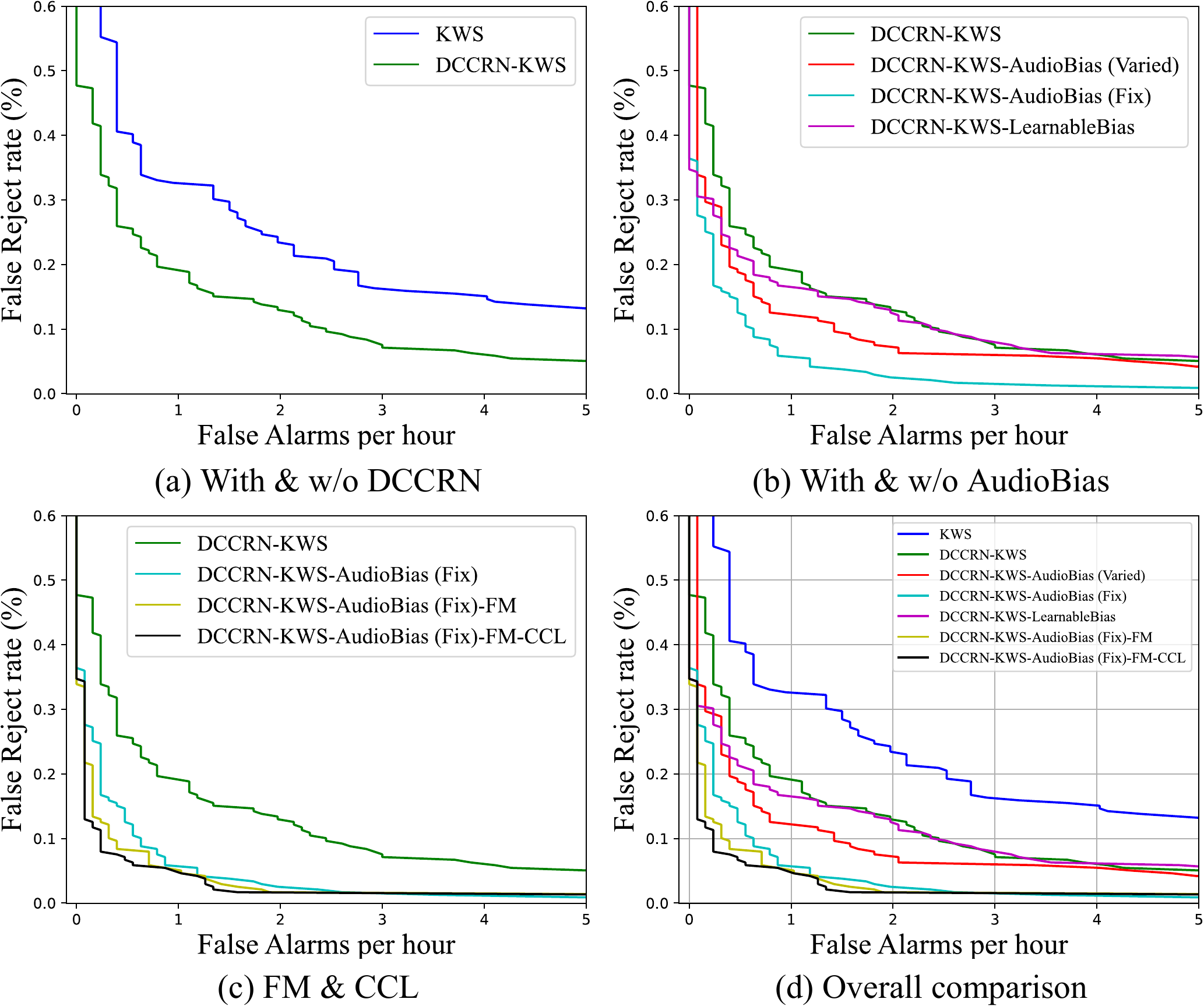}
\vspace{-0.2cm}
\caption{Results of comparison models and ablation of the proposed model on the internal dataset. FM: Feature Merge; CCL: Complex Context Linear.}
\vspace{-0.6cm}
\label{fig:dingdang}
\end{figure}

\vspace{-0.2cm}
As shown in Fig~\ref{fig:dingdang}, ablation is conducted to evaluate the effectiveness of different model components of our system, including a) DCCRN module, b) Audio context bias module (Fix / not Fix), c) Feature merge module (FM), and d) Complex context linear module (CCL). To show the effectiveness of audio context bias, we replace the bias embedding with a learnable bias embedding as a sanity check. For this learnable bias embedding, we first initialize a group of parameters as the bias embedding, then we send the bias embedding into the DCCRN-KWS module to update the value with the model training.
Performance is measured by plotting the receiver operating curve (ROC), which calculates the false reject (FR) rate per false alarm (FA) rate. The lower the FR per FA rate, the better the system.



It can be seen from Fig~\ref{fig:dingdang}(a) that after applying the DCCRN module, the performance becomes obviously better. This indicates that the use of an explicit speech enhancement module under multi-task learning framework is helpful to assist the KWS module to improve performance. From Fig~\ref{fig:dingdang}(b), adding audio context bias module yields a lower false reject rate when the false alarm rate kept the same with the system without the bias module, which shows the effectiveness of keyword audio embedding. We notice that fixed keyword audio list (Fix) performs better than varied list during training. This is probably because the model can be more easily trained and optimized better with a fixed list during training.

By comparing the learnable embedding with the embedding extracted by ECAPA-TDNN, we can see that the embedding extracted by keyword audio list is more effective than the learnable embedding. We can explain this phenomenon from several aspects. First, the embedding extracted from keyword audio contains more structure information of keywords than the learnable embedding. Second, we extract the bias embedding from real audio samples, which is easy to calculate the similarity with the DCCRN-KWS model compared with the learnable embedding. Furthermore, as shown in Fig~\ref{fig:dingdang}(c), when the feature merge module is applied, the performance is further improved. This is because the feature merge module can emphasize the discrimination of the keyword part as plotted in Fig~\ref{fig:feature_merge}, which can help the KWS module to distinguish the keyword from others. Finally, when the complex context linear module is adopted, our system achieves the best performance. This indicates that combining the context information and bias embedding can better discriminate the keyword. In summary, the overall comparison on all methods is shown in Fig~\ref{fig:dingdang}(d).


\begin{figure}[t]
\centering
\includegraphics[width=0.75\linewidth]{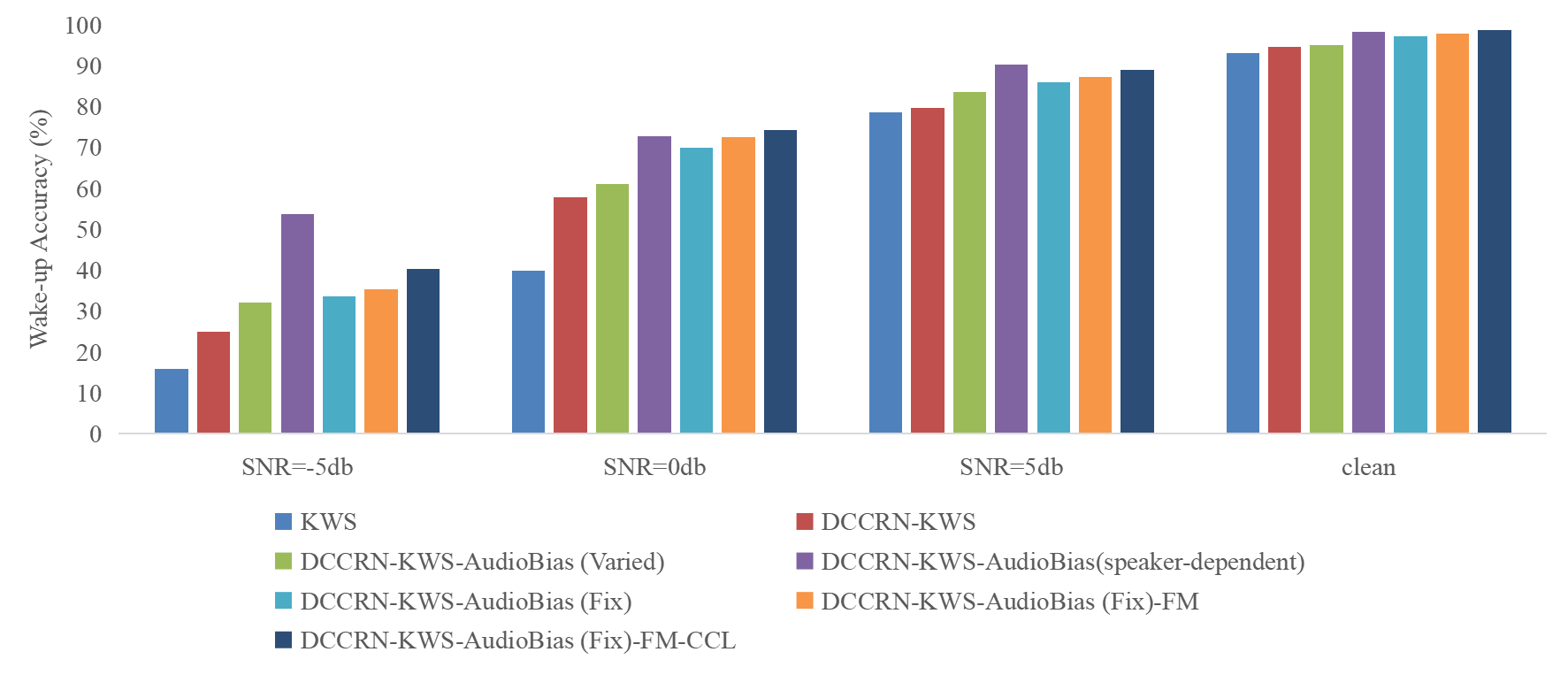}
\vspace{-0.4cm}
\caption{Results of different models and ablation of the proposed model on public HIMIYA dataset.} 
\label{fig:himiya}
\vspace{-0.6cm}
\end{figure}
\vspace{-0.4cm}
\subsubsection{Results on HIMIYA dataset}
\vspace{-0.2cm}
We further evaluate our models trained on the HIMIYA dataset. Fig~\ref{fig:himiya} shows the KWS performance measured by wake-up accuracy under the setup that up to one time false alarm triggered in 10 hours’ exposure to continuous noisy speech. In low SNR, high SNR and even clean scenarios, our contributions show their effectiveness according to the results. Please note that when speaker-dependent audio samples are applied as the audio list, the performance is better than the use of the speaker-independent audio list. This is because, for the speaker-dependent audio list, the extracted embedding contains both keyword information and speaker information. Furthermore, we notice that the improvements achieved from the proposed methods are less on the HIMIYA dataset than the internal dataset. The main reason is that the duration of the enrollment audio in the  HIMIYA dataset is relatively shorter than that in the internal dataset, while empirically a longer bias keyword will lead to a more robust keyword embedding. 
\vspace{-0.2cm}
\subsection{Inference Time}
\vspace{-0.1cm}
\label{sec:print}
As shown in Table~\ref{tab:rtf}, we evaluate the RTF and CPU usage on RK3326 Cortex-A35@1.5GHz low-resource platform. Inference precision is set at float32 with ONNX runtime engine. We also implement a comparison system, named DCCRN-KWS-joint-train in Table~\ref{tab:rtf}. In this comparison system, we jointly train a cascaded DCCRN-KWS system with the two losses, where the enhanced speech by DCCRN is fed to the DTC-based KWS system. We can see that the CPU usage and RTF for this system are not acceptable. In contrast, the proposed DCCRN-KWS has reasonable CPU usage and RTF. Importantly, based on the DCCRN-KWS framework, the components we introduced yield only a small increase in RTF. 

    \begin{table}[!h]
    \vspace{-15pt}
    \centering
    \footnotesize
    \setlength\tabcolsep{5pt}
    \vspace{0.1cm}
    \caption{RTF and CPU usage on low resource platform.} 
    \vspace{-0.2cm}
    \begin{tabular}{lcc}
    \toprule 
    Model & CPU Usage (\%) & RTF  \\ 
    \midrule 
    KWS  & \textbf{16} &  \textbf{0.27} \\ 
    DCCRN-KWS & 35 & 0.67 \\ 
    ~ + AudioBias & 36 & 0.69 \\ 
    ~ ~ + FM & 36 & 0.70 \\ 
    ~ ~ ~ + CCL & 38 & 0.72 \\ 
    DCCRN-KWS-joint-train & 94 & 2.85 \\ 
    \bottomrule
    \label{tab:rtf}
    \end{tabular}
    \vspace{-20pt}
    \end{table}
\vspace{-0.3cm}
\section{Conclusion}
\vspace{-0.1cm}
\label{sec:page}
This paper introduces a front-end and back-end integration framework for KWS in noisy conditions, named DCCRN-KWS. Specifically, we shape the integration in a multi-task learning manner, while DCCRN is adopted for speech enhancement and its encoder output is coupled with the KWS model for keyword spotting. Based on this MTL architecture, we propose an audio bias module that aims to better learn the discrimination between keywords and non-keywords. Feature merge and complex context linear modules are also introduced to strengthen such discrimination and to effectively leverage contextual information respectively. Experiments on two datasets show the effectiveness of the proposed approach. 

\bibliographystyle{IEEEtran}
\bibliography{mybib}

\begin{thebibliography}{10}
\providecommand{\url}[1]{#1}
\csname url@samestyle\endcsname
\providecommand{\newblock}{\relax}
\providecommand{\bibinfo}[2]{#2}
\providecommand{\BIBentrySTDinterwordspacing}{\spaceskip=0pt\relax}
\providecommand{\BIBentryALTinterwordstretchfactor}{4}
\providecommand{\BIBentryALTinterwordspacing}{\spaceskip=\fontdimen2\font plus
\BIBentryALTinterwordstretchfactor\fontdimen3\font minus
  \fontdimen4\font\relax}
\providecommand{\BIBforeignlanguage}[2]{{%
\expandafter\ifx\csname l@#1\endcsname\relax
\typeout{** WARNING: IEEEtran.bst: No hyphenation pattern has been}%
\typeout{** loaded for the language `#1'. Using the pattern for}%
\typeout{** the default language instead.}%
\else
\language=\csname l@#1\endcsname
\fi
#2}}
\providecommand{\BIBdecl}{\relax}
\BIBdecl

\bibitem{miller2007rapid}
D.~R. Miller, M.~Kleber, C.-L. Kao, O.~Kimball, T.~Colthurst, S.~A. Lowe, R.~M.
  Schwartz, and H.~Gish, ``Rapid and accurate spoken term detection,'' in
  \emph{Interspeech}, 2007.

\bibitem{wang2006computational}
D.~Wang and G.~J. Brown, \emph{Computational auditory scene analysis:
  Principles, algorithms, and applications}.\hskip 1em plus 0.5em minus
  0.4em\relax Wiley-IEEE press, 2006.

\bibitem{williamson2015complex}
D.~S. Williamson, Y.~Wang, and D.~Wang, ``Complex ratio masking for monaural
  speech separation,'' \emph{IEEE/ACM transactions on audio, speech, and
  language processing}, vol.~24, no.~3, pp. 483--492, 2015.

\bibitem{tan2018convolutional}
K.~Tan and D.~Wang, ``A convolutional recurrent neural network for real-time
  speech enhancement.'' in \emph{Interspeech}, 2018, pp. 3229--3233.

\bibitem{li2021simultaneous}
A.~Li, W.~Liu, X.~Luo, G.~Yu, C.~Zheng, and X.~Li, ``{A Simultaneous Denoising
  and Dereverberation Framework with Target Decoupling.}'' in
  \emph{Interspeech}, 2021, pp. 2801--2805.

\bibitem{hu2020dccrn}
Y.~Hu, Y.~Liu, S.~Lv, M.~Xing, S.~Zhang, Y.~Fu, J.~Wu, B.~Zhang, and L.~Xie,
  ``D{CCRN}: Deep complex convolution recurrent network for phase-aware speech
  enhancement.'' in \emph{Interspeech}, 2020, pp. 2472--2476.

\bibitem{reddy2020interspeech}
C.~K. Reddy, V.~Gopal, R.~Cutler, E.~Beyrami, R.~Cheng, H.~Dubey,
  S.~Matusevych, R.~Aichner, A.~Aazami, S.~Braun, P.~Rana, S.~Srinivasan, and
  J.~Gehrke, ``{The INTERSPEECH 2020 Deep Noise Suppression Challenge:
  Datasets, Subjective Testing Framework, and Challenge Results},'' in
  \emph{Interspeech}, 2020, pp. 2492--2496.

\bibitem{reddy2021interspeech}
C.~K. Reddy, H.~Dubey, K.~Koishida, A.~Nair, V.~Gopal, R.~Cutler, S.~Braun,
  H.~Gamper, R.~Aichner, and S.~Srinivasan, ``{INTERSPEECH 2021 Deep Noise
  Suppression Challenge.}'' in \emph{Interspeech}, 2021, pp. 2796--2800.

\bibitem{o2021conformer}
T.~O'Malley, A.~Narayanan, Q.~Wang, A.~Park, J.~Walker, and N.~Howard, ``A
  conformer-based asr frontend for joint acoustic echo cancellation, speech
  enhancement and speech separation,'' in \emph{ASRU}.\hskip 1em plus 0.5em
  minus 0.4em\relax IEEE, 2021, pp. 304--311.

\bibitem{nian2022time}
Z.~Nian, J.~Du, Y.~T. Yeung, and R.~Wang, ``A time domain progressive learning
  approach with snr constriction for single-channel speech enhancement and
  recognition,'' in \emph{ICASSP}.\hskip 1em plus 0.5em minus 0.4em\relax IEEE,
  2022, pp. 6277--6281.

\bibitem{li2021espnet}
C.~Li, J.~Shi, W.~Zhang, A.~S. Subramanian, X.~Chang, N.~Kamo, M.~Hira,
  T.~Hayashi, C.~Boeddeker, Z.~Chen \emph{et~al.}, ``Espnet-se: end-to-end
  speech enhancement and separation toolkit designed for asr integration,'' in
  \emph{SLT}.\hskip 1em plus 0.5em minus 0.4em\relax IEEE, 2021, pp. 785--792.

\bibitem{kong2021multi}
Y.~Kong, J.~Wu, Q.~Wang, P.~Gao, W.~Zhuang, Y.~Wang, and L.~Xie,
  ``Multi-channel automatic speech recognition using deep complex unet,'' in
  \emph{SLT}.\hskip 1em plus 0.5em minus 0.4em\relax IEEE, 2021, pp. 104--110.

\bibitem{gu2020efficient}
Y.~Gu, Z.~Du, H.~Zhang, and X.~Zhang, ``An efficient joint training framework
  for robust small-footprint keyword spotting,'' in \emph{International
  Conference on Neural Information Processing}.\hskip 1em plus 0.5em minus
  0.4em\relax Springer, 2020, pp. 12--23.

\bibitem{yu2018text}
M.~Yu, X.~Ji, Y.~Gao, L.~Chen, J.~Chen, J.~Zheng, D.~Su, and D.~Yu,
  ``Text-dependent speech enhancement for small-footprint robust keyword
  detection.'' in \emph{Interspeech}, 2018, pp. 2613--2617.

\bibitem{shin2022learning}
H.-K. Shin, H.~Han, D.~Kim, S.-W. Chung, and H.-G. Kang, ``Learning audio-text
  agreement for open-vocabulary keyword spotting,'' \emph{arXiv preprint
  arXiv:2206.15400}, 2022.

\bibitem{hou2021npu}
J.~Hou, L.~Zhang, Y.~Fu, Q.~Wang, Z.~Yang, Q.~Shao, and L.~Xie, ``The npu
  system for the 2020 personalized voice trigger challenge,'' \emph{arXiv
  preprint arXiv:2102.13552}, 2021.

\bibitem{pundak2018deep}
G.~Pundak, T.~N. Sainath, R.~Prabhavalkar, A.~Kannan, and D.~Zhao, ``Deep
  context: end-to-end contextual speech recognition,'' in \emph{SLT}.\hskip 1em
  plus 0.5em minus 0.4em\relax IEEE, 2018, pp. 418--425.

\bibitem{desplanques2020ecapa}
B.~Desplanques, J.~Thienpondt, and K.~Demuynck, ``Ecapa-tdnn: Emphasized
  channel attention, propagation and aggregation in tdnn based speaker
  verification,'' \emph{arXiv preprint arXiv:2005.07143}, 2020.

\bibitem{speechbrain}
M.~Ravanelli, T.~Parcollet, P.~Plantinga, A.~Rouhe, S.~Cornell, L.~Lugosch,
  C.~Subakan, N.~Dawalatabad, A.~Heba, J.~Zhong, J.-C. Chou, S.-L. Yeh, S.-W.
  Fu, C.-F. Liao, E.~Rastorgueva, F.~Grondin, W.~Aris, H.~Na, Y.~Gao, R.~D.
  Mori, and Y.~Bengio, ``{SpeechBrain}: A general-purpose speech toolkit,''
  2021, arXiv:2106.04624.

\bibitem{luo2019conv}
Y.~Luo and N.~Mesgarani, ``Conv-tasnet: Surpassing ideal time--frequency
  magnitude masking for speech separation,'' \emph{IEEE/ACM transactions on
  audio, speech, and language processing}, vol.~27, no.~8, pp. 1256--1266,
  2019.

\bibitem{qin2020hi}
X.~Qin, H.~Bu, and M.~Li, ``Hi-mia: A far-field text-dependent speaker
  verification database and the baselines,'' in \emph{ICASSP}.\hskip 1em plus
  0.5em minus 0.4em\relax IEEE, 2020, pp. 7609--7613.

\bibitem{dong2018speech}
L.~Dong, S.~Xu, and B.~Xu, ``Speech-transformer: a no-recurrence
  sequence-to-sequence model for speech recognition,'' in \emph{ICASSP}.\hskip
  1em plus 0.5em minus 0.4em\relax IEEE, 2018, pp. 5884--5888.

\end{thebibliography}

\end{document}